%% file: main.tex
\documentclass[a4paper,12pt,authoryear,times]{elsarticle}
\usepackage[authoryear]{natbib}
\usepackage{graphicx} 
\usepackage{float}
\usepackage{algorithm}  
\usepackage{algpseudocode}
\usepackage{color}
\usepackage{setspace}

\usepackage[T2A, T1]{fontenc}			
\usepackage[utf8]{inputenc}			
\usepackage[russian, english]{babel}

\usepackage{amssymb}
\usepackage{amsmath}

\usepackage{ulem}

\usepackage[table]{xcolor}

\usepackage{amssymb}

\journal{ }

\begin{document}

\begin{frontmatter}

\title{A Deep Learning-Aided Approach for Estimating Field Permeability Map by Fusing Well Logs, Well Tests, and Seismic Data}

\author[SKT]{Grigoriy Shutov\cormark[1]}
\author[SKT]{Viktor Duplyakov\cormark[1]}
\author[SKT,TPU]{Shadfar Davoodi\cormark[2]}
\author[SKT]{Anton Morozov}
\author[SKT]{Dmitriy Popkov}
\author[IRMR]{Kirill Pavlenko}
\author[SKT]{Albert Vainshtein}
\author[GPN]{Viktor Kotezhekov}
\author[GPN]{Sergey Kaygorodov}
\author[GPN]{Boris Belozerov}
\author[GPN]{Mars M Khasanov}
\author[SKT]{Vladimir Vanovskiy\cormark[2]} 
\author[SKT]{Andrei Osiptsov}
\author[SKT,AIRI]{Evgeny Burnaev}

\cortext[1]{Equal contribution}
\cortext[2]{Corresponding author. \textit{Email addresses}: S.Davoodi@skoltech.ru (S. Davoodi); e.burnaev@skoltech.ru (E.Burnaev).} 

\affiliation[SKT]{organization={Skolkovo Institute of Science and Technology}, addressline={the territory of the Skolkovo Innovation Center, Bolshoy Boulevard, 30, bld. 1,}, city={Moscow}, postcode={143026}, country={Russian~Federation}}

\affiliation[TPU]{organization={School of Earth Sciences and Engineering, Tomsk Polytechnic University institution},addressline={30~Lenin~Avenue},  city={Tomsk}, postcode={634028}, country={Russian~Federation}}

\affiliation[IRMR]{organization={Independent Researcher}, city={Moscow}, country={Russian Federation}}
    
\affiliation[GPN]{organization={Gazprom~Neft~Group~of~Companies}, addressline={3~-5~Zoologichesky~Lane}, city={Saint~Petersburg}, postcode={190000}, country={Russian~Federation}}
    
\affiliation[AIRI]{organization={Autonomous Non-Profit Organization Artificial Intelligence Research Institute (AIRI)}, city={Moscow},country={Russian Federation}}

\begin{abstract}
Obtaining reliable permeability maps of oil reservoirs is crucial for building a robust and accurate reservoir simulation model and, therefore, designing effective recovery strategies. This problem, however, remains challenging, as it requires the integration of various data sources by experts from different disciplines. Moreover, there are no sources to provide direct information about the inter-well space. In this work, a new method based on the data-fusion approach is proposed for predicting two-dimensional permeability maps on the whole reservoir area. This method utilizes non-parametric regression with a custom kernel shape accounting for different data sources: well logs, well tests, and seismics.  A convolutional neural network is developed to process seismic data and then incorporate it with other sources. A multi-stage data fusion procedure helps to artificially increase the training dataset for the seismic interpretation model and finally to construct an adequate permeability map. The proposed methodology of permeability map construction from different sources was tested on a real oil reservoir located in Western Siberia. The results demonstrate that the developed map perfectly corresponds to the permeability estimations in the wells, and the inter-well space permeability predictions are considerably improved through the incorporation of the seismic data.
\end{abstract}

\begin{highlights}
\item Data fusion of oil reservoir permeability maps that integrates well and seismic data.
\item Convolutional neural network for predicting oil reservoir properties by seismic data.
\item Multi-stage fusion expands training data to enhance inter-well permeability prediction.
\end{highlights}
\end{frontmatter}

\input{1_intro}
\input{2_literature}
\input{3_database}
\input{4_methodology}
\input{5_results}
\input{6_conclusion}

\bibliographystyle{elsarticle-harv}
\bibliography{bibliography}

\appendix
\include{7_appendix}
\end{document}

%% file: 1_intro.tex
\newpage

\section{Introduction}
\label{sec:introduction}

Reservoir simulations are the basis of robust forecasting of reservoir production performance and designing proper recovery strategies \citep{shen2021comparative, hajibolouri2024permeability}. Among the reservoir properties, permeability plays one of the most crucial roles in reservoir simulation models, since it quantifies a rock formation's ability to transmit fluids through the reservoir formation \citep{olatunji2014hybrid, jahanbakhsh2016gas, chang2023new}. Permeability maps or cubes provide a spatial representation of the transmission capability of petroleum reservoirs, which could be beneficial for optimal well placement and designing effective recovery strategies of oil and gas fields under development \citep{evans2019mapping, zhang2023estimating, kanin2024combined}. 

Different methods have been applied for the determination of reservoir permeability. Core analysis is the only direct technique used for permeability measurement, while indirect methods include well log and seismic data analysis. \citep{xue2022reservoir, otchere2021novel, matinkia2023prediction}. Except for the seismic, all these methods suffer from data source locality, as well as measurement errors that vary across different reservoirs \cite{matinkia2023prediction, pan2022optimized, abdulaziz2022influences}. The integration of such data into a reservoir simulation model usually requires the work of a team of specialists who interpret raw data from the standpoint of their respective specializations \cite{cannon2024reservoir}.  Each of them, while creating the component parts of the object model, often does not see the full picture. In particular, seismic specialists create multiple geobodies, whereas geophysicists, independently of each other, characterize reservoir properties. In addition, well logs and well tests provide valuable information about permeability around wells; however, permeability in the inter-well space often comes with high inherent uncertainty \cite{ganguli2023reservoir}.  The procedure of creating an adequate reservoir simulation model is quite complex and time-intensive, typically taking months, depending on the characteristics of the field \cite{ng2021smart}. 

To overcome these difficulties, the present study develops a new method for permeability map prediction. Data fusion techniques, which industries often utilize when integrating multiple data sources \cite{castanedo2013}, serve as the basis for the proposed method. As a result, it combines the data of different sources and localities and then fuses them into a single permeability map.

The present method is based on non-parametric regression with a special kernel shape accounting for different data locality and reliability. As for the data being fused, the interpretations of well logs and well tests, and global seismic RMS amplitude cubes are used. To process the latter into permeability values, a convolutional neural network (CNN) model is developed. The multi-stage fusion procedure helps to artificially extend the training dataset for the seismic interpretation model by using grid points with high estimated fidelity of permeability prediction by well log and test data fusion. It is worth noting that although fusion is used for developing 2D permeability maps, the method proposed can be extended to other fields, such as conductivity or porosity, and in 3D.

The developed data fusion method is evaluated on the real oil reservoir in Western Siberia. The quality of the method is estimated based on how well the fused permeability map corresponds to well measurements that were unseen to the method’s algorithm. As a result,  the present method not only fuses the permeability maps that correspond to well measurements—permeability estimations derived from well logs and well tests—but also estimates the regions where such measurements are relevant. In the inter-well space, on the other hand, the developed algorithm automatically relies on seismic data rather than the well measurements. Therefore, it integrates data from various sources and localities in a robust way. 

In summary, the main  contributions of the present study  are:
\begin{enumerate}
    \item An automatic kernel regression-based method with interpretable hyperparameters that combines well logs and well test interpretations together into a single permeability map.
    \item A convolutional neural network (CNN) model that predicts the target properties based on RMS amplitude cubes. 
    \item Multi-stage fusion procedure that artificially extends the amount of data used for training the seismic data interpretation model and allows one to obtain an adequate permeability prediction in the inter-well space. 
\end{enumerate}

%% file: 2_literature.tex
\section{Literature review}
\label{sec:literature}

The goal of data fusion in the method proposed in this work is to ensure that each source complements the other to ultimately achieve adequate reservoir characterization. This section contains a review of the definition of data fusion, its use in various fields, an evaluation of its applicability in reservoir characterization, and a comparison of it with similar geostatistical problems.

Typically, data fusion is utilized in industries for integrating multiple data sources \cite{castanedo2013}. Representative fusion problems include but are not limited to wireless sensor network systems \cite{gavel2021data}, image processing \cite{solsona2017new}, radar systems, object tracking \cite{wang2024deep}, target detection and identification \cite{liu2021robust}, network intrusion detection \cite{ayantayo2023network}, situation assessment \cite{pengfei2023anti}, life science \cite{smilde2022multiblock}. For example, data fusion is highly utilized in urban domains, where problematic data is addressed while enhancing the data reliability and extracting knowledge from multiple data sources \cite{lau2019survey}. When it comes to reservoir modeling, the fusion of various geophysical data, such as seismic, electrical, magnetic, gravity, radiometric, and thermal, is used to increase the reliability of the conclusions obtained by certain geophysical exploration methods. Fusion is mandatorily used in well logging and often in ore and engineering geology.
In this regard, data fusion methods, such as geostatistical techniques, have also found their applications in the development of spatial maps for reducing the uncertainties associated with sample data limitations to estimate a reservoir property. These methods are also popular for their prominent efficiency in the analysis of petrophysical data to map the values of subsurface properties in areas with unsampled data \cite{evans2019mapping}. Classical geostatistical methods consider the spatial dependence between observations (either from single or various data source types) to predict values at unsampled locations. One of the most used stochastic interpolation methods is kriging of different kinds, the exact choice of which depends on the stochastic properties of the random field and degrees of stationary assumptions \cite{kanin2024combined}. Regardless of the stochastic methods, deterministic-based methods, such as the inverse distance weighting (IDW) that estimates unknown values based on a simple mathematical weighting function \cite{sadeghi2024inverse}.  The kernel regression approach used in the present study can be considered as an extended version of the IDW method to arbitrary weighting functions, depending also on the different types of interpolated data (similar to cokriging in this aspect). Kernel regression can be considered a strong alternative to cokriging, particularly for fusion problems involving data sources with different locality. This is due to the fact that the kernel regression has superior capability for setting adjustments to account for variability in data locality.
The existing review made on applications of data fusion techniques in the oil and gas industry confirms its great potential in addressing different problems in this field. For instance, Li et al. (2019) \cite{li2019fusing} employed a machine-learning-based fusion approach to combine multiple decomposed-frequency seismic attributes for accurate prediction of sand thickness of sand bodies using an SVM model.  Fusing of seismic attributes showed improved correlation with the sand thickness compared to individual systemic attributes \cite{li2019fusing}. To estimate the field porosity map, Xu et al. (2015) applied a co-kriging method to fuse two different data sources, namely seismic attributes with well-log-derived porosity data collected from the Blackfoot field based in Canada. The proposed cokriging system achieved an improved-isolation porosity map across the studied field with low variance in estimation errors. They highlighted that developed porosity mapping techniques can be effective for predicting the target parameter in areas with sparse wells \cite{xu2015porosity}. Evans et al. (2019) \cite{evans2019mapping} have applied simple geostatistical methods, simple and ordinary kriging techniques, to estimate permeability and porosity maps in the Jubilee oilfield located in Ghana. Applying geostatistical kriging enabled the authors to estimate the targeted petrophysical parameter values in unsampled locations far from the sampled areas \cite{evans2019mapping}. To estimate the spatial distribution of permeability across a gas reservoir situated within the Khangiran field, Hosseini et al. geostatistical modeled the permeability, applying techniques such as ordinary kriging and sequential Gaussian simulation based on well-log-derived permeability values. It was highlighted that the kriging method provided a credible spatial permeability map, enabling estimation of the target properties of unsampled reservoir parts. Although the outcome of proposed geostatistical methods provided spatial-based maps of permeability parameters, their being solely dependent on a single petrophysical data source (e.g., core and petrophysical logs) limits their reliability. It is worth noting that core and log data are representative of petrophysical properties in near-wellbore areas and prone to high uncertainty associated with interpretation and measurement errors. Therefore, the integration of different data sources is recommended to improve the reliability of spatial mapping of petrophysical parameters and alleviate the locality issue with data. To some extent, the most similar approach to geostatistical interpolation is data fusion, but the key difference between them is that interpolation takes only one source as input data and then propagates its properties in the inter-well space, while fusion predicts target values in the inter-well space based on multiple data sources.  

In conclusion, data fusion is a broad topic that can be extensively categorized in different ways and is applicable in various data-intensive fields. As for reservoir characterization, few works have approached fusing multiple sources, and even fewer works have considered permeability maps as the reservoir characterization target, to the best of our knowledge. To some extent, the most similar problem to data fusion is geostatistical interpolation, but the key difference between them is that the interpolation takes only one source as input data and then propagates its properties in the inter-well space, while fusion takes into account multiple sources and predicts target values in the inter-well space based on these sources. Hence, the proposed method is novel in the sense that it fuses three relevant data sources and asserts permeability as a target.

%% file: 3_database.tex
\section{Data overview}
\label{sec:data}

\begin{figure}[h!!!]
    \centering
    \includegraphics[width=\linewidth]{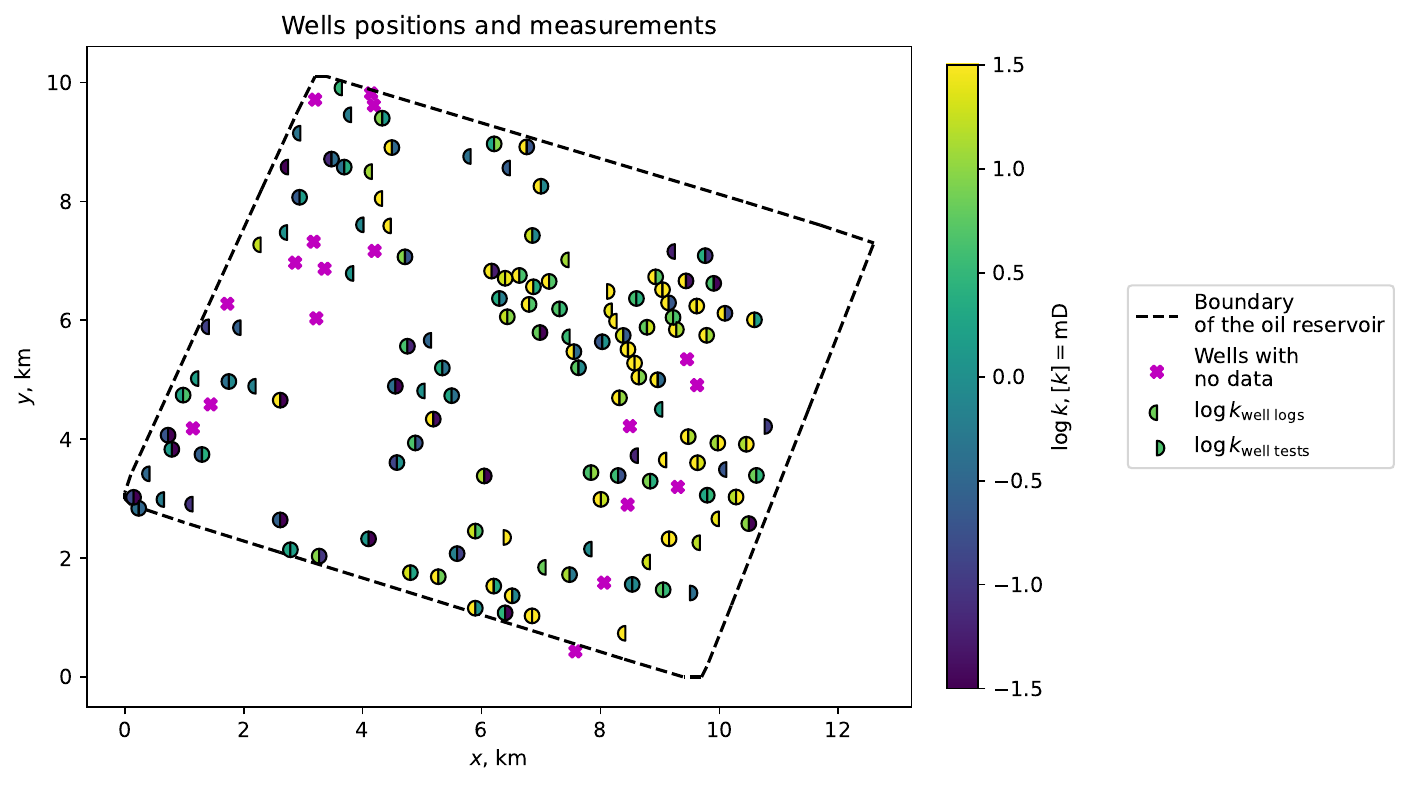}
    \caption{Well location map presenting the positions of the wells and the availability of permeability measurements in those wells, as well as the interpreted values. The oil reservoir boundary depicts the area in which the target permeability map is aimed to be fused.}
    \label{fig:data_availiability}
\end{figure}

To perform this study, real data from a part of the oilfield in Western Siberia with several productive formations were compiled. In the stratigraphic plan, the reservoirs belong to the Cretaceous deposits of the Upper Vartovskaya suite, composed of frequent and uneven intercalations of mudstones with sandstones and siltstones. As for the geological and physical characteristics of the studied formations, effective oil-saturated thicknesses vary from 0.4 to around 60 m, averaging 13 m for the deposit; the true vertical depth of productive formations varies from 2400 to 2600 m; and the formation temperature is in the range of 88–90 °C. At the same time, only the most productive formation and a separate area with 147 wells (see Fig. \ref{fig:data_availiability}) are considered below due to data availability (see Figure \ref{fig:data_availiability}).

The following data sources were used during the study:

\begin{itemize}
    \item  well log data: permeability ($K_{wl}$) and net pay ($H_{wl}$);
    \item  well test data -- effective permeability from well test ($K_{wt}$), net pay ($H_{wt}$), and date and type of the survey;
    \item seismic cube data: root mean square amplitudes (RMS);
    \item special core analysis (SCAL) filtration data for each rock type: oil and water relative permeability curves;
    \item well coordinates \& horizons;
    \item formation fluids properties (oil \& water viscosity).
\end{itemize}

The choice of reservoir ought to be driven by the least tendency to anisotropy of properties; therefore, it is recommended to use a field with a single productive formation. The selected field is a more challenging case for such analysis because of the operation of wells in multiple reservoirs, so the data are required to be pre-processed to obtain separate values for each of the formations.

In the reservoir examined, 88 wells possess both well-log and well-test measurements; 37 wells provide only well-log data, 4 wells supply only well-test data, and 18 wells lack measurements altogether. The spatial distribution of these wells is shown in Figure \ref{fig:data_availiability}.

\subsection{Data sources}

Well-log permeability, $K_{wl}$, is obtained from porosity logs via calculated petrophysical dependencies. Such transformation contains several uncertainties associated with the following sources:

\begin{itemize}
    \item the error of the method itself for determining porosity from logs correlates with the error of measuring tool: interpretation results are highly dependent on the quality and duration of measurements \citep{Ma2011};
    \item the spread of the porosity function $\phi = f(log(K_{wl}))$ is quite high due to the specificity of core studies and interpretation, as the core sample characteristics are measured at surface conditions instead of reservoir ones. Overall, it leads to vague permeability values \citep{Shi2017,Hall1958}.
\end{itemize}

As for the build-up well tests, such permeability, $K_{wt}$, is closest in its physical meaning to that calculated directly in the reservoir simulation model and is an integral filtration characteristic of the well drainage area \citep{stepiko2018eng}. However, switching from effective to absolute permeability is also needed since the latter corresponds to a rock characteristic rather than its combination with the composition of formation fluids \citep{SATTER201629}. The following formula is, therefore, utilized to convert effective permeability to absolute one:
\begin{equation} \label{eqn:eff_to_abs_k}
    K_{abs} = \frac{K_{wt}}{\mu_{liq}(\frac{kr_o(S_w)}{\mu_o} + \frac{kr_w(S_w)}{\mu_w})},
\end{equation}
where $K_{wt}$ – effective permeability from well test, $\mu_{liq}$, $\mu_{o}$, $\mu_{w}$ – liquid, oil and water viscosity, respectively; $kr_o(S_w)$, $kr_w(S_w)$ – relative oil and water permeabilities, respectively, which are found from rock type relative phase diagrams. 

\begin{figure}[h!!!]
\centering\includegraphics[width=1\textwidth]{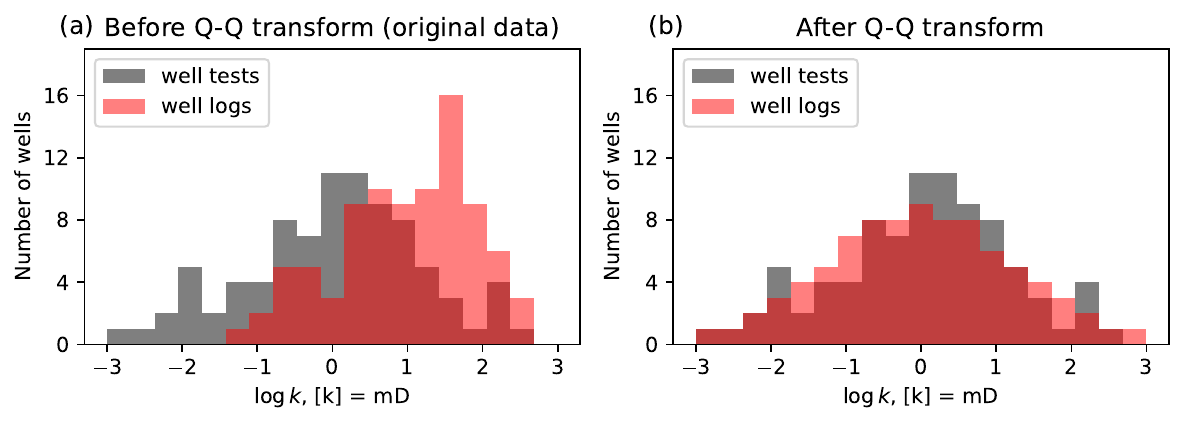}
\caption{Comparison of permeability distributions from well log and well test data before (a) and after (b) Q-Q transformation.}
\label{fig:q-q}
\end{figure}

As seen in Figure \ref{fig:data_availiability}, the well log-based permeability measurements show a higher concentration of zones with low permeability as indicated by darker colors (purple and blue), though a few high-permeability zones (green to yellow) are observed in the wells considered. The well test-based permeability measurements shown in Figure \ref{fig:data_availiability} display almost a similar trend, with a higher concentration of wells with low permeability values. The comparison of the histograms (of the well test and well log permeability values in Figure \ref{fig:q-q}a demonstrates that there is variability in permeability estimates depending on the measuring method since well logs often provide higher resolution of permeability values at the wellbore level, while well tests offer a broader representation of permeability values at certain drainage areas around the wells. This comparison confirms the importance of using multiple data sources to assess reservoir permeability rather than solely relying on a single source.

During methodology validation on the field data, a so-called problem <<bull-eyes>> at the well positions on the fused permeability cubes was encountered, which is described in Section \ref{sec:results}. The origin of the problem lies in the frequent discrepancies between well logs and well tests interpreted permeabilities. Consequently, this results in either a local increase or decrease in values near the well since logging data is much more reliable in a small radius (see Section \ref{sec:methodology}). To solve this problem, To solve this problem, the logging data were transformed via a quantile-quantile (Q-Q) transformation (Figure \ref{fig:q-q}b). The transformation of the logging data can be performed either to log-normalize the distribution, as the permeability distribution in the real field is log-normal \citep{law1944}, or to match the well tests target distribution.

Figure \ref{fig:q-q} compares the distribution of the permeability log estimated by well log data (pink) with those calculated based on well test data (blue), highlighting the discrepancy degree in the permeability estimated from the two data sources before and after the Q-Q transformation. As observed in Figure \ref{fig:q-q}a, the permeability distribution derived from well log data does not align with that of the well test data. The distribution of well test permeability appears to have a higher density at the lower values, while the well log-based permeability displays a narrower distribution with a larger density at higher values. A comparison of the two distributions after applying the Q-Q transformation (Figure \ref{fig:q-q}b) demonstrates that the well test permeability distribution matches that estimated by well log data, suggesting that the Q-Q transformation effectively reduced the discrepancies observed between the permeability cube estimates derived from the well test and well log data. 

%% file: 4_methodology.tex
\section{Methodology}

\begin{figure}[h!!!]
    \centering
    \includegraphics[width=1\linewidth]{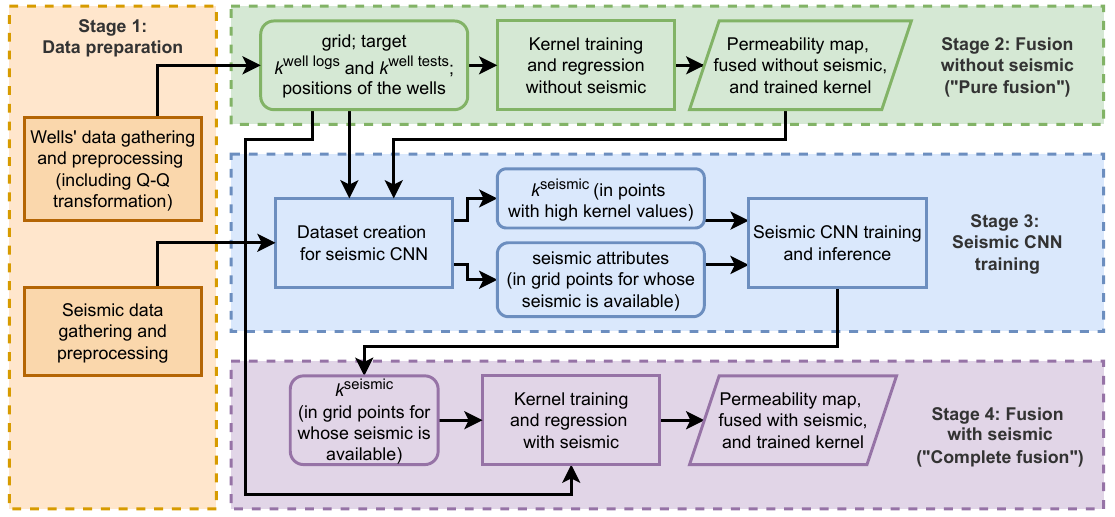}
    \caption{Workflow diagram illustrating the methodology applied for mapping the spatial distribution of permeability. The process consists of four stages: (1) data collection and preprocessing, (2) kernel-regression based "pure fusion," (3) CNN model training for predicting permeability from seismic, and (4) kernel-regression based complete fusion of well log, well-test, and seismic data, estimating a refined final permeability.}
    \label{fig:whole_workflow}
\end{figure}

The methodology applied to reliably map the spatial distribution of permeability across the considered area of the studied field is diagrammatically displayed in Figure \ref{fig:whole_workflow}. The developed methodology consists of four subsequent stages. In the first stage, the required well log, well test, and seismic data are collected from the field under study and preprocessed (see Section \ref{sec:data}). In the second stage, wells interpreted permeabilities by well logs and well tests are used in kernel regression to fuse permeability map (see Section \ref{sec:methodology}). The fusion process at this stage is called "pure fusion," in which seismic data are not used. Next, the purely fused permeability map from the previous stage, combined with preprocessed seismic data, are compiled into a dataset to train a CNN model for predicting the global target map at high kernel-value (or high fidelity) points (see Section \ref{sec:seismic}). Finally, the seismic CNN-predicted permeability, along with those from the well test and well log sources, are used to train a new kernel and refine the target fusion. This final stage is called “complete fusion,” which integrates all three sources considered in our study to estimate the permeability map.

It is important to note that although the present study applies this fusion method to develop 2D permeability maps, the approach is readily extendable to predict 3D cubes of permeability and other petrophysical properties, such as porosity.

\subsection{Kernel regression}
\label{sec:methodology}

The purpose of the data fusion approach is to obtain the most accurate initial absolute permeability map for the considered area. So, all the information sources (well logs, well tests, and seismics) for the absolute permeability have to be combined, taking into account their various levels of uncertainty and the degree of confidence depending on the distance from the well. The procedure is validated through a <<leave-one-out cross validation>> (LOOCV) strategy, where each well value is used for validation by comparing it with the model results without this well.

The following approach is proposed to precisely establish the exact relationship between data sources (well logs, well tests, and seismics). The logarithm of the absolute permeability is taken as the target variable, based on the assumption that the spread of the target variable may span several orders of magnitude \citep{Roding2020}. The initial approximation for the permeability map is obtained using the Nadaraya-Watson kernel regression \citep{nadaraya1964estimating, watson1964smooth} estimator for target permeability $K$, which represents a weighted average over all points $\Vec{r_i}$ (measured from each well) surrounding the considered point $\Vec{r}$, as given in Equation \ref{nadaraya_reg}:
\begin{equation} \label{nadaraya_reg}
    K(\Vec{r}) = \frac{\sum_{i=1}^N \mathcal{K}(\Vec{r}-\Vec{r}_i) k_i}{\sum_{i=1}^N \mathcal{K}(\Vec{r}-\Vec{r}_i)},
\end{equation}
where $\vec{r} = (x, y)$ is a radius-vector, $K(\Vec{r})$ is the estimated permeability map, and $k_i$ is the permeability from $i$th source. The kernel $\mathcal{K}(r)$ accounts for the varying locality and reliability of the three data sources used: well test $\mathcal{K}_{\mathrm{wt}}$, well log $\mathcal{K}_{\mathrm{wl}}$, and seismic $\mathcal{K}_{\mathrm{seismic}}$ kernels:
\begin{align}
    \label{bi_kernel}
        &\mathcal{K}_{\mathrm{wt}}(\vec{r}-\vec{r}_{\mathrm{w}}) = (|\vec{r}-\vec{r}_{\mathrm{w}}|/r_d)^{\alpha}e^{-(|\vec{r}-\vec{r}_{\mathrm{w}}|/r_d)^\beta}\\
        &\mathcal{K}_{\mathrm{wl}}(\vec{r}-\vec{r}_{\mathrm{w}}) = \gamma e^{-(|\vec{r}-\vec{r}_{\mathrm{w}}|/r_g)^\delta}\label{eq:well_log_kernel}\\
        &\mathcal{K}_{\mathrm{seismic}}(\vec{r}-\vec{r}_{\mathrm{w}}) =\mathcal{K}_{\mathrm{seismic}} = w_{s} \label{eq: seismic kernel}
\end{align}
where $\vec{r}_{\mathrm{w}}$ is well position, $\alpha$ is locality index, $\beta$ is relative well log radial exponent, $\gamma$ is low-fidelity factor of well log data, $\delta$ is relative well test radial exponent, $r_d$ is well test radius (i.e., drainage radius), $r_g$ is geological permeability correlation radius, and $w_s$ is seismic weight value (constant). These kernel parameters $\alpha$, $\beta$, $\gamma$, $\delta$, $r_d$, $r_g$, and $w_s$ will be optimized, and the optimization will be described further. 

\begin{figure}[h!!!]
\centering
\includegraphics[width=0.9\textwidth]{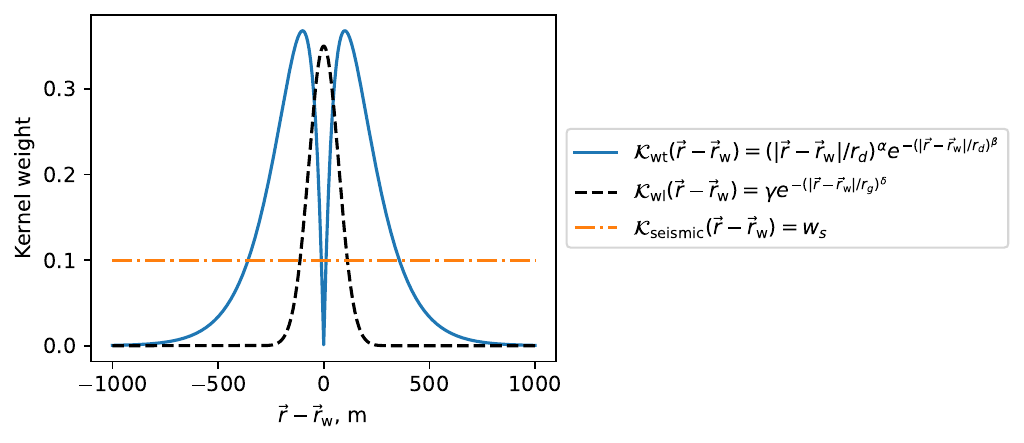}
\caption{Kernel weights as a function of distance from well for different data sources. $\mathcal{K}_{\mathrm{wt}}$, $\mathcal{K}_{\mathrm{wl}}$ and $\mathcal{K}_{\mathrm{seismic}}$ represent kernels of permeability derived from well test, well log, and seismic data respectively for a single well at $\vec{r}_w$.}
\label{fig:kernel}
\end{figure}

Figure \ref{fig:kernel} illustrates the dependence of the kernel weights of the target for the well test, well log, and seismic data on the distance to a well. According to the kernel weights illustrated, the permeability derived from well logs reaches the highest impact in the immediate vicinity of the well ($\mathcal{K}_{\mathrm{wl}}$), while that from well test data primarily affects the target distribution within the approximate drainage radius ($r_d$ in $\mathcal{K}_{\mathrm{wt}}$). Meanwhile, the permeability derived from seismic data refines inter-well space independent of distance, meaning its kernel weight remains constant throughout the field ($\mathcal{K}_{\mathrm{seismic}}$). Thus, kernel weights properly determine data reliability for each source considered at a given distance from the well. It should be noted that the relationships presented in Figure \ref{fig:kernel} correspond to a single well at $\vec{r}_w$ as an example to visualize the kernel weights for the target distribution from multiple sources considered.

Using such a kernel, a permeability map $K(\vec{r})$ is introduced, enabling the estimation of the target value at a given distance by incorporating data from the available $N$ wells, as presented in Equation \eqref{eq:fusion}:
\begin{equation} \label{eq:fusion}
    K(\vec{r}) = \frac{\sum_{i=1}^N \mathcal{K}_{\mathrm{wt}}(\vec{r}-\vec{r}_{i}) k_{i}^{\mathrm{wt}} + \sum_{i=1}^N \mathcal{K}_{\mathrm{wl}}(\vec{r}-\vec{r}_{i}) k_{i}^{\mathrm{wl}} + \mathcal{K}_{\mathrm{seismic}} K^{\mathrm{seismic}}(\vec{r})} {\sum_{i=1}^N \left[\mathcal{K}_{\mathrm{wt}}(\vec{r}-\vec{r}_{i}) + \mathcal{K}_{\mathrm{wl}}(\vec{r}-\vec{r}_{i})\right] +  \mathcal{K}_{\mathrm{seismic}}},
\end{equation}
where $\vec{r}_i$ is a position of $i$th well, $k_{i}^{\mathrm{wt}}$ is interpreted well tests permeability, $k_{i}^{\mathrm{wl}}$ is interpreted well logs permeability, and $K^{\mathrm{seismic}}(\vec{r})$ is permeability derived from seismic. It is important to note that $K^{\mathrm{seismic}}(\vec{r})$ is not readily available for any reservoir. To obtain this parameter, a predictive seismic model needs to be developed, which will be described further in Section \ref{sec:seismic}. Therefore, at this stage, the fusion is conducted without seismic data by assuming that $\mathcal{K}_{\mathrm{seismic}} \equiv 0$.

The oil reservoir can be represented on a grid, which is a set of points $\Omega = \{ \vec{r}_j \}_{j=1}^{N_{\mathrm{grid}}}$. The reservoir studied in this work has a grid of 7962 points. However, the grid is chosen by user, and the algorithm can work with various grids. Figure \ref{fig:data_availiability} demonstrates the boundary that surrounds all these points in the grid. When the permeability map is fused, the formula given in Equation \eqref{eq:fusion} is applied for computing the target, $k$, in each point of the grid, $\vec{r}_j$, to obtain the permeability map, $K = \{ k_j \}_{j=1}^{N_{\mathrm{grid}}}$. 

During the kernel optimization process the fused permeability values in the wells need to be checked and verified. Hence, the inverse task occurs: the permeability map, $K$, is known, and the permeability $k^{\mathrm{w}}_i$ in the well at location $\vec{r}_i$, estimated from the map, needs to be determined. One way to compute this is to use Equation \eqref{eq:fusion}. However, this approach may lead to an issue: after optimization, the map might exhibit reasonable permeability values at the wells' positions, $\vec{r}_i$, while displaying absolutely unreasonable values elsewhere. To mitigate this problem, it is essential to consider all map values around the well, $\vec{r}_i$. Thus, permeability at the wells is computed using the kernel approach, as defined in Equation \eqref{eq:in_well}.
\begin{align}
\label{eq:in_well}
\begin{split}
    &k^{\mathrm{w}}_i = \frac{\sum_{j}^{N_{\mathrm{grid}}} K(\vec{r}_j) \mathcal{K}(\vec{r}_j - \vec{r}_i)}{\sum_{j}^{N_{\mathrm{grid}}} \mathcal{K}(\vec{r}_j - \vec{r}_i)} \\
    &\mathcal{K}(\vec{r}_j - \vec{r}_i) = \exp \left(-\frac{|\vec{r}_j - \vec{r}_i|}{r_{\mathrm{dr}}} \right)
\end{split}
\end{align}
where $k^{\mathrm{w}}_i$ denotes the permeability in the $i$th well and $r_{\mathrm{dr}}$ is the drainage radius, which is set to 250 m in the present study. This means that the permeability in the well, $k^{\mathrm{w}}_i$, is a weighted sum of the permeability map values, $k_j$, from the permeability map $K$. The weighting scheme assigns higher weights to the points near the well and lower weights to those farther away. Such aggregated permeability estimates are referred to as synthetic well tests.

The following optimization approach (see Algorithm 1) is applied to determine the optimal values of kernel constants ($\alpha$, $\beta$, $\gamma$, $\delta$, $r_d$, $r_g$, and $w_s$) for the field case considered using a leave-one-out-cross-validation (LOOCV) technique. This process is performed by applying differential evolution \citep{storn1997differential}. An issue that may arise during this process is that the optimizer could converge to a solution where all the target values on the map become uniform, which leads to negligible differences between the scenarios with and without the excluded well during LOOCV. To overcome this problem, regularization is introduced into the objective functions. Also the kernel parameters are sought in geologically reasonable ranges (for example $r_d$ in the range of the possibile drainage radii). These parameters in general can be set by any geologist working with the oilfield and such possibility demonstrates the interpretability of the selected kernel approach.  Thus, the objective function to be minimized is defined as in Equation \eqref{eq:opt_func}.   

\begin{algorithm}
    \label{alg:kernel_opt}
    \caption{Kernel training via LOOCV. \\ \textbf{Input:} wells positions $\mathcal{R}=\{\vec{r}_i\}_{i=1}^N$; target values in wells by well logs $\{k^{\mathrm{wl}}_i\}_{i=1}^N$ and well tests $\{k^{\mathrm{wt}}_i\}_{i=1}^N$; grid $\Omega = \{\vec{r}_j\}_{j=1}^{N_{\mathrm{grid}}}$; initial guess kernel parameters $\alpha$, $\beta$, $\gamma$, $\delta$, $r_d$, $r_g$, $w_s$; optimizer parameters (number of optimization steps $n_{\mathrm{iter}}.$, boundaries of kernel parameters etc.). \\ \textbf{Output:} optimized parameters $\alpha$, $\beta$, $\gamma$, $\delta$, $r_d$, $r_g$, $w_s$.}
    \begin{algorithmic}[1]
    \For{iteration in $1, \ldots, n_{\mathrm{iter}}$}
    \For{well $i$ in $1, \ldots, N$}
    \State Exclude the $i$th well and obtain the new set of wells $\hat{\mathcal{R}}=\mathcal{R} \backslash \{\vec{r}_i\}$, which is a set of all wells except the well at $\vec{r}_i$;
    \State Using Equation \eqref{eq:fusion} together with current kernel parameters and wells' sets $\mathcal{R}$ and $\hat{\mathcal{R}}$, fuse two permeability maps $K$ and $\hat{K}$, respectively;
    \State Using Equation \eqref{eq:in_well} and permeability maps $K$ and $\hat{K}$, compute the <<synthetic well-tests>> $k_i^{\mathrm{w}}$ and $\hat{k}_i^{\mathrm{w}}$ in the $i$th well;
    \EndFor
    \State Given <<synthetic well-tests>> $k=\{k_i^{\mathrm{w}}\}_{i=1}^N$ and $\hat{k}=\{\hat{k}_i^{\mathrm{w}}\}_{i=1}^N$, compute the metric \eqref{eq:opt_func};
    \State Update the parameters $\alpha$, $\beta$, $\gamma$, $\delta$, $r_d$, $r_g$, $w_s$ using optimizer.
    \EndFor
    \end{algorithmic}
\end{algorithm}

\begin{align} \label{eq:opt_func}
\begin{split}
    &f = 1 - R^2 \left( k, \hat{k} \right) + c_1l_1 + c_2l_2,\\
    &l_1 = \frac{1}{\sqrt{2}} \sqrt{\left| \mathrm{PDF}(k) - \mathrm{PDF}(\hat{k}) \right|} \\
    &l_2 = \frac{|(\max k - \min k) - (\max \hat{k} - \min \hat{k})|}{\max(k, \hat{k})}
\end{split}
\end{align}
where $R^2$ represents the coefficient of determination between the true and predicted values synthetic well-test values; $l_1$ denotes the difference between the true and predicted values distributions, normalized between 0 and 1; $l_2$ represents the difference between the true and predicted $max - min$ target differences, normalized; $c_1, c_2$ are the regularization coefficients.

Due to its shape depicted on Figure \ref{fig:kernel}, one can say that the kernel values are always higher close to wells and lower away from wells. Indeed, its values reflect the confidence of the fused data. We discuss this matter in Section \ref{sec:pure fusion}.

For brevity, the methodology is described only for permeability; however, the same approach can be applied to fuse hydraulic conductivity maps or porosity maps.

\subsection{Automatic seismic analysis}
\label{sec:seismic}

Due to its huge size, seismic data contain a vast set of information, ranging from geobody location to specific attributes such as sweetness, chaos, and reflection intensity. Correlations between seismic data and petrophysical properties are extensively explored through machine learning methods in previous studies \citep{Iturraran2014, Anifowose2019, Wei2019}. However, one of the primary challenges associated with this approach is the limited size of the training dataset, which is typically constrained by well-logging data. This limitation hinders the training of sufficiently expressive neural networks, which often necessitates the use of shallow models. In this study, a method is proposed to integrate seismic data automatically with kernel regression predictions for permeability (or conductivity) maps. The proposed approach fuses seismic and well-logging data while seismic data is processed using a deep convolutional network.

Seismic data is obtained through reflected waves in the modification of the common depth point. Root mean square amplitudes (RMS) serve as a fundamental seismic attribute, which has no direct correlation with permeability. For the reservoir considered in this work, seismic data consists of  416 inlines and 515 crosslines with a discretization step of 25 meters. The vertical discretization step is set at 2 ms. 

For each grid point $\vec{r} = (x, y)$ on the 2D permeability map, the corresponding $z$-coordinate is determined, which represents the center of the associated formation. For each triplet $(x, y, z)$, a cube of the fixed length $\delta_s$ with the center in $(x, y, z)$ is constructed, and the RMS data for a given point are only read from the corresponding cube. Each cube has a size of (9, 9, 46). The objective is to match these RMS cubes with the permeability (conductivity) values at the grid point $\vec{r} = (x, y)$.

\begin{figure}[h!!!]
    \centering
    \includegraphics[width=1\linewidth]{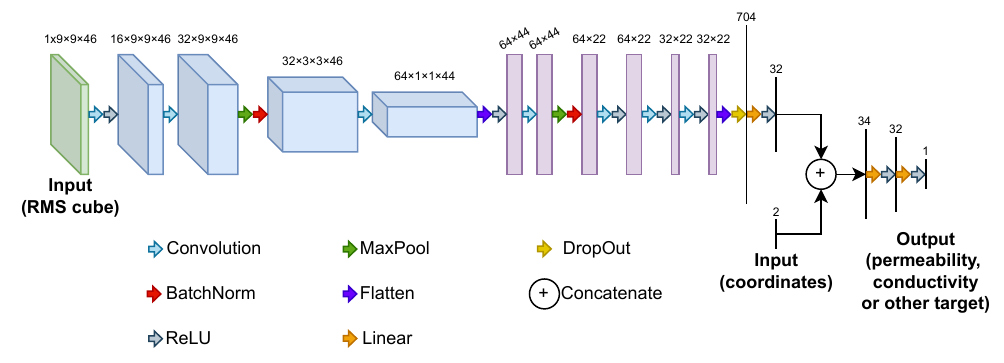}
    \caption{Detailed architecture of the seismic CNN model. The model receives two inputs: seismic RMS cubes surrounding specified points and coordinates of those points. The CNN outputs predictions of the target parameter, permeability. Before the first flattening operation, the network architecture consists of several layers of three-dimensional convolution, max pooling, batch normalization, and rectified linear activation functions (ReLU), indicated by blue blocks. After flattening, the data underwent sequential one-dimensional convolutional layers (purple blocks).}
    \label{fig:architecture}
\end{figure}

To achieve this, a 3D deep convolutional neural network (3D CNN) is trained to take seismic data as input and predict permeability. The CNN architecture is presented in Figure \ref{fig:architecture}). In practice, well-logging and well-testing target data are insufficient for effectively training a neural network. Due to the limited size of the training dataset (typically no more than a few hundred points), deep neural networks tend to overfit because of the large number of trainable parameters. The key insight that enables the training of such models effectively is \textit{training data expansion}. To address the issue of small sample size, the training dataset is extended using kernel regression. Instead of relying solely on the permeability directly obtained from well-logging and well-testing, high-confidence predictions from kernel regression are incorporated as some artificial wells. The confidence of predicted value is quantified by the value of the kernel at the corresponding point. Therefore, points with kernel values above $P_{50}$ percentile were included in the training dataset. It should be noted that such percentile threshold is a hyperparameter that is chosen by a user.

By including predicted permeability values with high confidence, the training dataset is expanded from a few hundred points to several thousand. Qualitatively, this approach enables the inclusion of the points that are located in the vicinity of the wells.

The CNN model was trained using RMS data extracted from seismic measurements. This approach enabled the model to learn the correlation between RMS seismic data and the permeability targets near well locations. Subsequently, predictions were extended across the entire mapped area, including regions where kernel regression data (derived from logs and well tests) were unavailable or unreliable. Validation of the proposed approach was conducted by removing random patches from the training dataset and assessing model performance on these excluded segments.

Although prediction accuracy near wells was observed to be slightly lower compared to kernel regression methods, the CNN model demonstrated robust predictive performance in regions located far from wells, thereby maintaining stable prediction
capabilities across areas lacking direct well data. Once trained, the CNN allowed for inference of seismic-based target values $K^{\mathrm{seismic}}(\vec{r})$ at every grid point ($K^{\text{seismic}} =\{ k^{\text{seismic}}_j \}_{j=1}^{N_{\text{grid}}}$), facilitating the subsequent training of new kernels that incorporated seismic information (see Equation \eqref{sec:seismic}).

%% file: 5_results.tex
\section{Results and discussion}
\label{sec:results}
This section describes the results of "pure fusion", in which only permeability maps derived from well-test and well-log data are fused, and "complete fusion", which incorporates data from all three sources: well tests, well logs, and seismic. 

\subsection{Fusion without seismic and Q-Q transformation}
\label{sec:pure fusion}
\begin{figure}[h!!!]

\centering\includegraphics[width=\textwidth]{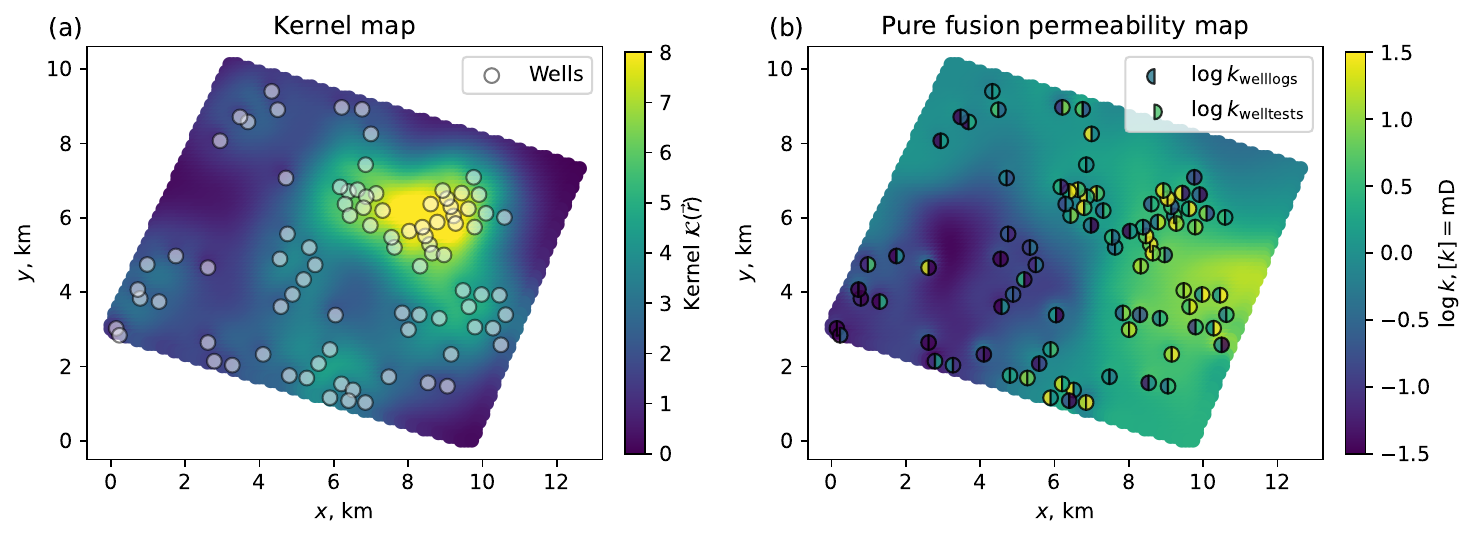}
\caption{a) Kernel map for the given reservoir, showing the kernel values trained using well log- and well test-derived permeability values, and b) permeability fused using only well log and well test data ("pure fusion").}
\label{fig:pure_fusion}
\end{figure}

The kernel training algorithm (see Algorithm 1) was first applied to train the kernel using the well log and well test data, and the trained kernel was then used to fuse permeability maps from these two sources (this step corresponds to Stage 2 of the proposed workflow in Figure \ref{fig:whole_workflow}). Figure \ref{fig:pure_fusion} shows the trained kernel and fused permeability maps generated based on this pure fusion approach.

As shown in Figure \ref{fig:pure_fusion}a, the kernel exhibits higher values near the wells and lower values in areas farther from the wells. The closer a point is to a well, the more the kernel values align with the actual values, leading to greater confidence in the predicted values. Conversely, in areas distant from the wells, where no data on the target value is available except for seismic data, the predicted values are expected to have higher uncertainty. Thus, the kernel values, $\mathcal{K}(\vec{r})$, serve as a reliable indicator of the confidence in the predicted target value at a given point $\vec{r}$. 

In the fused permeability map shown in Figure \ref{fig:pure_fusion}b, it can be observed that the map values correspond to those in the wells. For synthetic well tests (see Eq.\eqref{eq:in_well}), a coefficient of determination $R^2=0.96$ was achieved. It can also be seen that the most conductive zone in the fused permeability map is associated with the high-density number of wells.

\begin{figure}[h!!!]
\centering\includegraphics[width=0.99\linewidth]{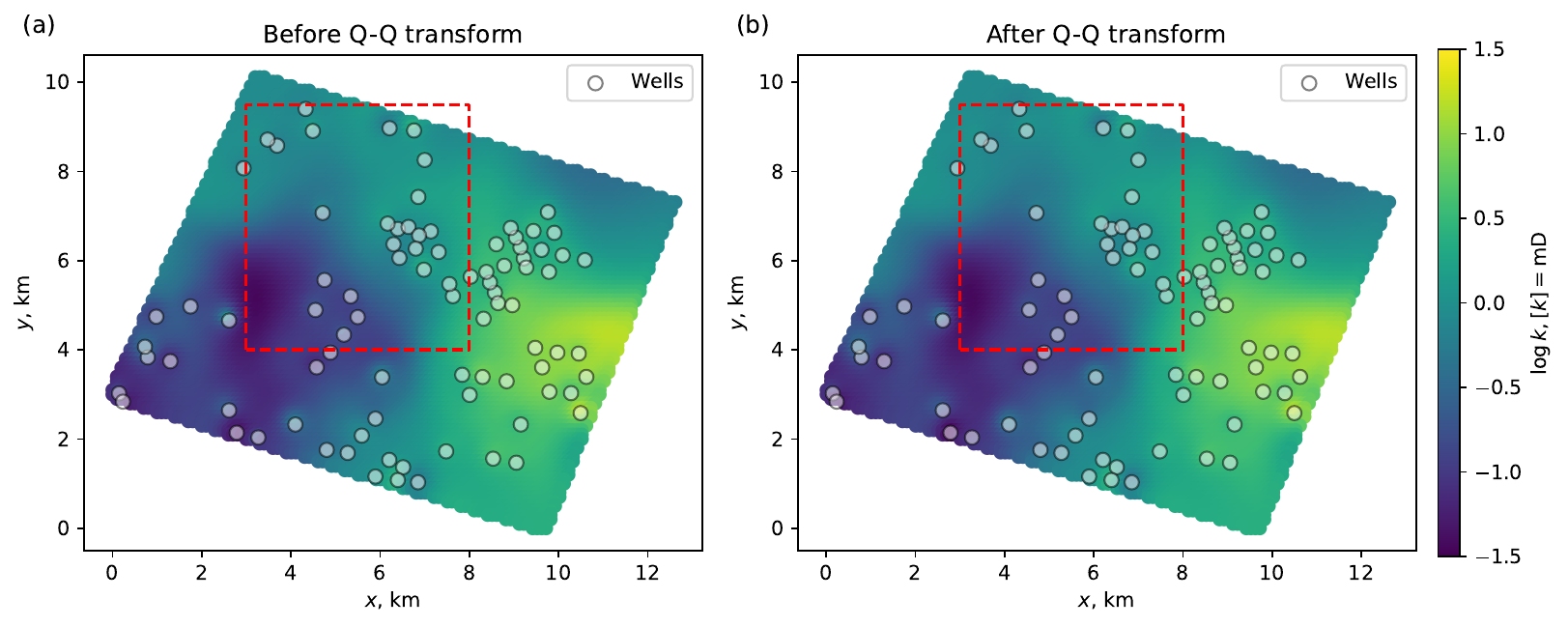}
\centering\includegraphics[width=0.65\linewidth]{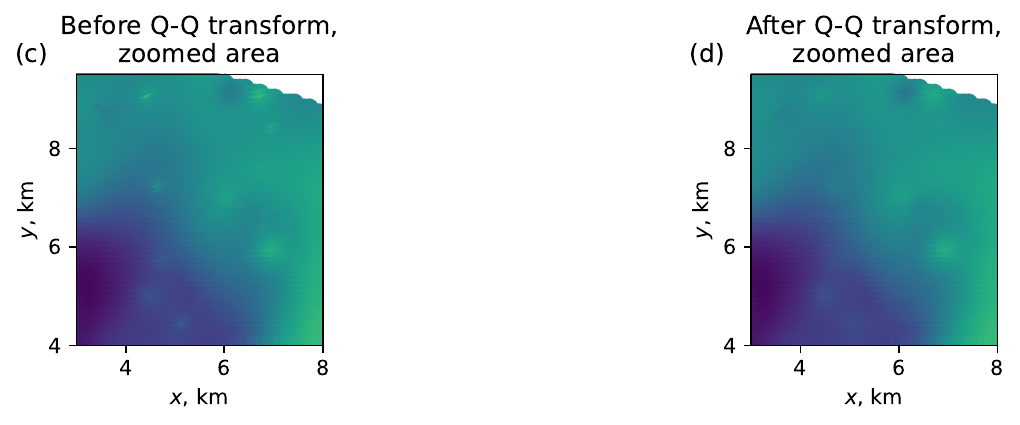}
\caption{Permeability maps generated using the “pure fusion” approach: (a) without and (b) with applying the Q-Q transformation. The red dashed box highlights a zoomed-in region shown in (c) and (d), respectively. The zoomed-in region prior to transformation (c) reveals a pronounced "bull-eye" effect, which is effectively mitigated when the Q-Q transformation is applied (d).}
\label{fig:q-q_result}
\end{figure}

The influence of the Q-Q transformation on the permeability map generated based on the “pure fusion” method is presented in Figure \ref{fig:q-q_result}. This transformation effectively mitigates the issue of "punched points," where permeability values at well locations significantly diverge from adjacent grid cells. Although it does not entirely remove the "bull-eye" effect—characterized by noticeable contrasts between near-well and inter-well permeability—the transformation reduces its severity. It facilitates a smoother transition between high-permeability zones near the wells and the surrounding reservoir, improving the spatial coherence of the permeability map. Overall, the Q-Q transformation adjusted the overestimated permeability values derived from well test data, bringing them into closer agreement with those derived from well log measurements.

\subsection{Seismic CNN training and fusion with seismic}
This subsection presents the results obtained from Stage 3 of the methodology (see Fig.\ref{fig:pure_fusion}), in which a seismic 3DCNN was trained using the extracted kernel weights and the permeability values derived from the "pure fusion" approach. During training, kernel values corresponding to high-confidence predictions from the kernel regression were employed as the training dataset, whereas cells with low kernel weights were excluded and used solely for inference.

\begin{figure}[h!!!]
\centering\includegraphics[width=\textwidth]{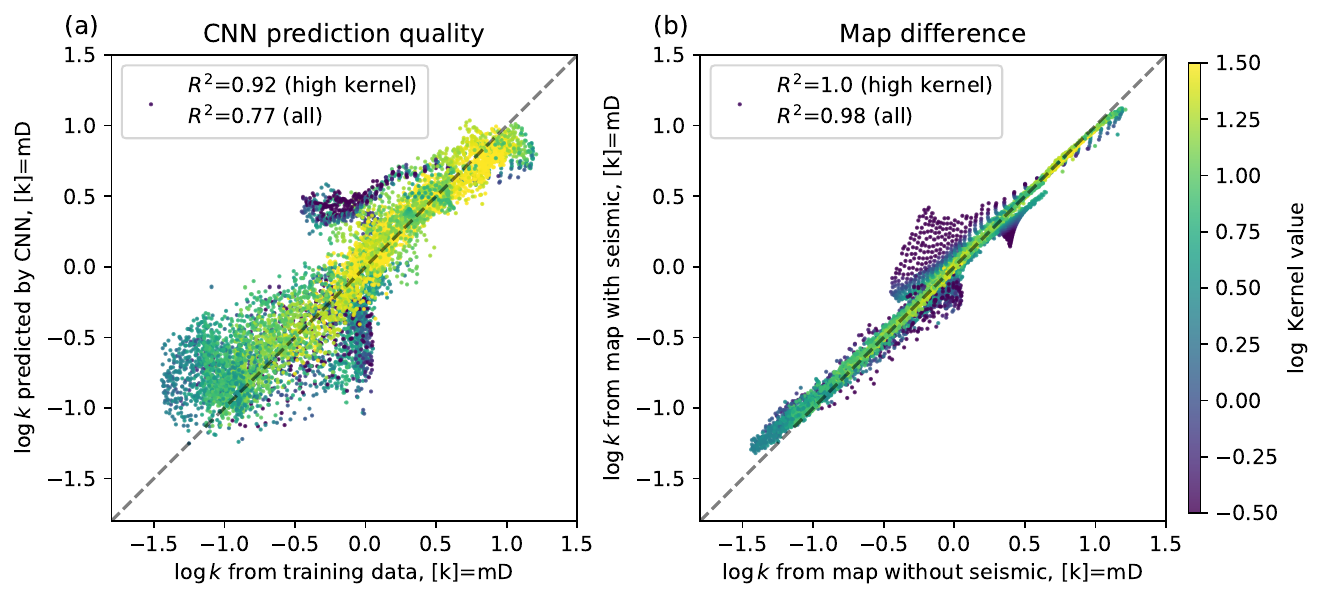}
    \caption{(a) Comparison of log-permeability predictions from the trained seismic 3DCNN versus those obtained from the "pure fusion" approach. (b) Comparison of log-permeability values derived from data fusion incorporating seismic information versus those obtained without seismic data.}
\label{fig:learning_curve}
\end{figure}

Figure \ref{fig:learning_curve}a cross plots the permeability values predicted by the seismic 3DCNN versus those obtained from the "pure fusion" map. Figure \ref{fig:learning_curve}b depicts the cross plots comparing permeability values obtained from data fusion that incorporates seismic information ("complete fusion")  with those from fusion excluding seismic input, "pure fusion."

As observed in Figure \ref{fig:learning_curve}a, a high $R^2$ value of 0.92 was achieved for log permeability predictions generated by the trained seismic-3DCNN with respect to high kernel values, while a lower $R^2$ value (0.77) was obtained when permeability predictions with low-kernel values were considered. This outcome indicates that, due to the lower confidence of "pure fusion" permeability estimates in regions of low kernel values, the predictions made by the seismic-3DCNN in areas of high kernel weight can be considered more reliable.

From Figure \ref{fig:learning_curve}b, it can be observed that the inclusion of the seismic data introduces new information for the points with low initial prediction confidence (far from the known ones near the wells). This result highlights the importance of the seismic data as a clarifying source for the points in the inter-well space.

\begin{figure}[h!!!]
\centering\includegraphics[width=0.6\linewidth]{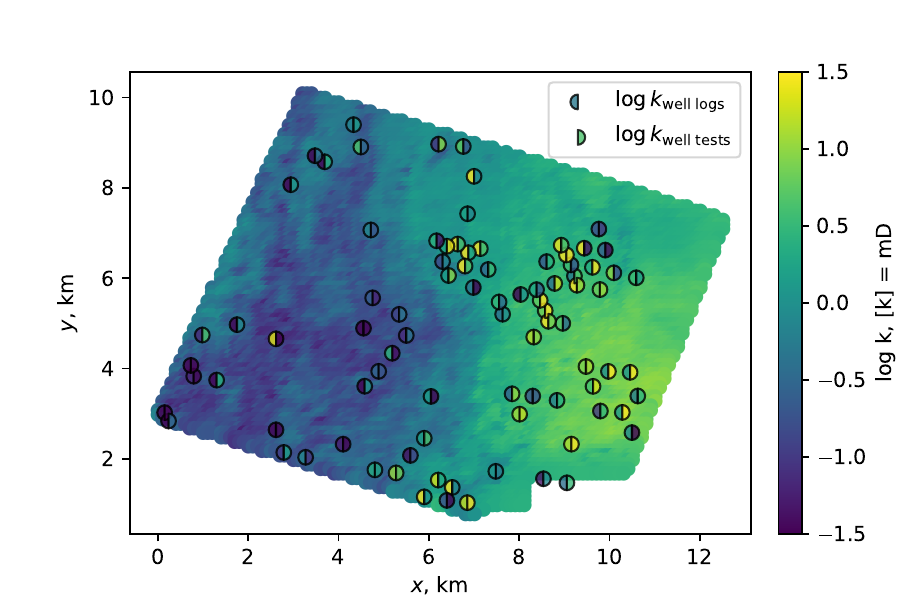}
\caption{Permeability map predicted by the trained 3DCNN using solely seismic attributes. The color scale denotes $\log k$; green-to-yellow colors indicate higher permeability values, whereas blue-to-purple colors represent lower permeability. Although the data from the wells, well-log- and well-test-interpreted permeabilities are not used in the generation of this map, they are demonstrated for the sake of comparison.}
\label{fig:seismic_weight_1}
\end{figure}

Following the seismic 3DCNN training, a permeability map was developed exclusively based on the seismic data. This map is presented in Figure \ref{fig:seismic_weight_1}. When compared with the reference map presented in Figure \ref{fig:pure_fusion}, the seismic-based permeability map predictions matched pretty well in the vicinity of the wells and supplied supplementary information within the inter-well space. 

As displayed in Figure \ref{fig:seismic_weight_1}, the permeability predictions obtained from seismic data alone are relatively noisy. Therefore, those predictions should be judiciously used, primarily within the inter-well spaces. This limitation was addressed through the automatic selection of the seismic weight in the data-fusion procedure (see Eq. 7).

\begin{figure}[h!!!]

\centering\includegraphics[width=\textwidth]{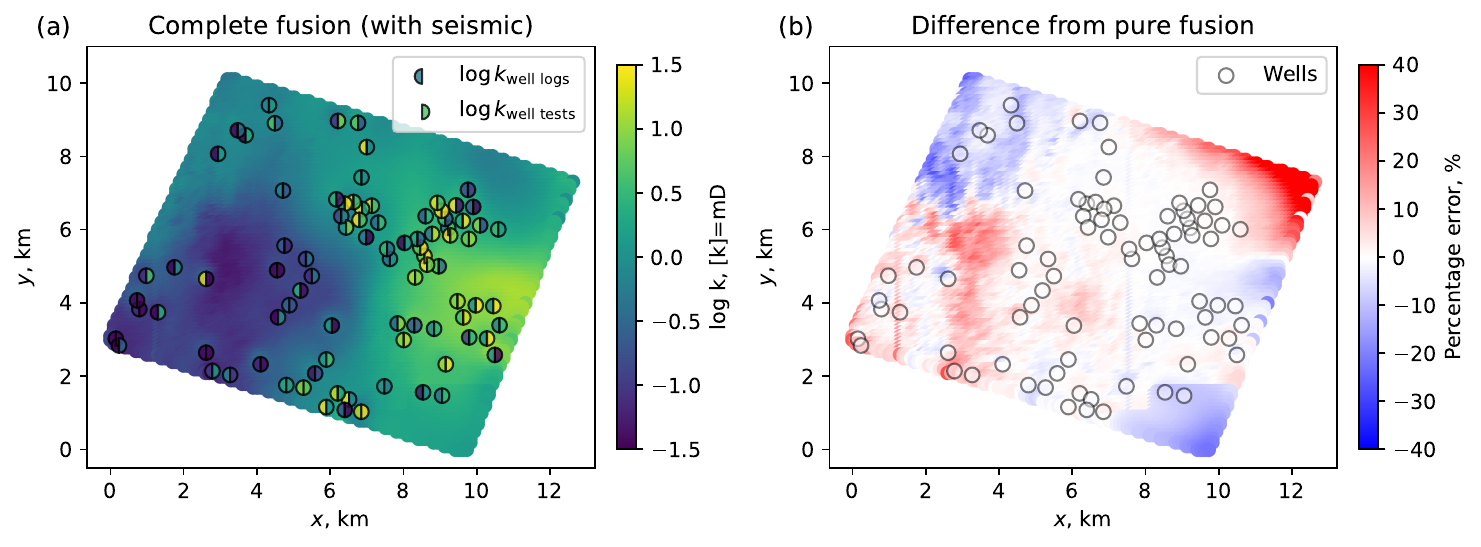}
\caption{Comparison of permeability maps: a) Permeability map from "complete fusion" of the well test, well logs, and seismic data; b) Computed difference map of the "complete fusion" map and the "pure fusion" map, which was presented in Figure 7a.}
\label{fig:complete_fusion}
\end{figure}

After estimating the seismic permeability by the 3DCNN model $K^{\mathrm{seismic}}(\vec{r})$, a new kernel was optimized with the training procedure described in Algorithm 1. During this optimization, the seismic weight $w_S$ was allowed to be a non-zero value so that the resulting field permeability map represented the fusion of data from all three sources: well log, well test, and the seismic (this step corresponds to Stage 4 of the proposed workflow in Figure \ref{fig:whole_workflow}). The fused map of permeability is displayed in Figure \ref{fig:complete_fusion}a.

By comparing the permeability map presented in Figure \ref{fig:complete_fusion}a for the "complete fusion" result with the map for the "pure fusion" result (see Figure \ref{fig:pure_fusion}b), it can be seen that additional spatial detail was introduced in regions distant from wells by incorporating the seismic data, while the noise level remained low as a result of the selected optimal seismic weight.

For a better understanding of the seismic influence, Figure \ref{fig:complete_fusion}b presents a comparison between the “complete fusion” and “pure fusion” maps by displaying the percentage difference, which is computed using the following formula. 
\begin{equation} \label{eq:percentage error1}
   \text{Percentage error }(\vec{r}) = 100 \cdot\frac{K_{\text{complete fusion}}(\vec{r}) - K_{\text{pure fusion}}(\vec{r})}{K_{\text{pure fusion}}(\vec{r})}
\end{equation}

By considering the difference percentage map in Figure \ref{fig:complete_fusion}b together with the results presented earlier in Figure \ref{fig:learning_curve}b, it can be concluded that the incorporation of seismic cubes predominantly influences regions located far from the wells, where permeability estimations from well logs and well tests are not available. These results clearly confirm the behavior anticipated from the proposed kernel approach.

\subsection{Ablation study}
\label{sec:ablation}
An ablation study was performed to validate the final permeability map by excluding wells characterized by either low or high permeability values. After the wells were excluded from the data, the whole workflow (see Fig. \ref{fig:whole_workflow}) was repeated completely. 

\begin{figure}[h!!!]
\centering\includegraphics[width=1.0\textwidth]{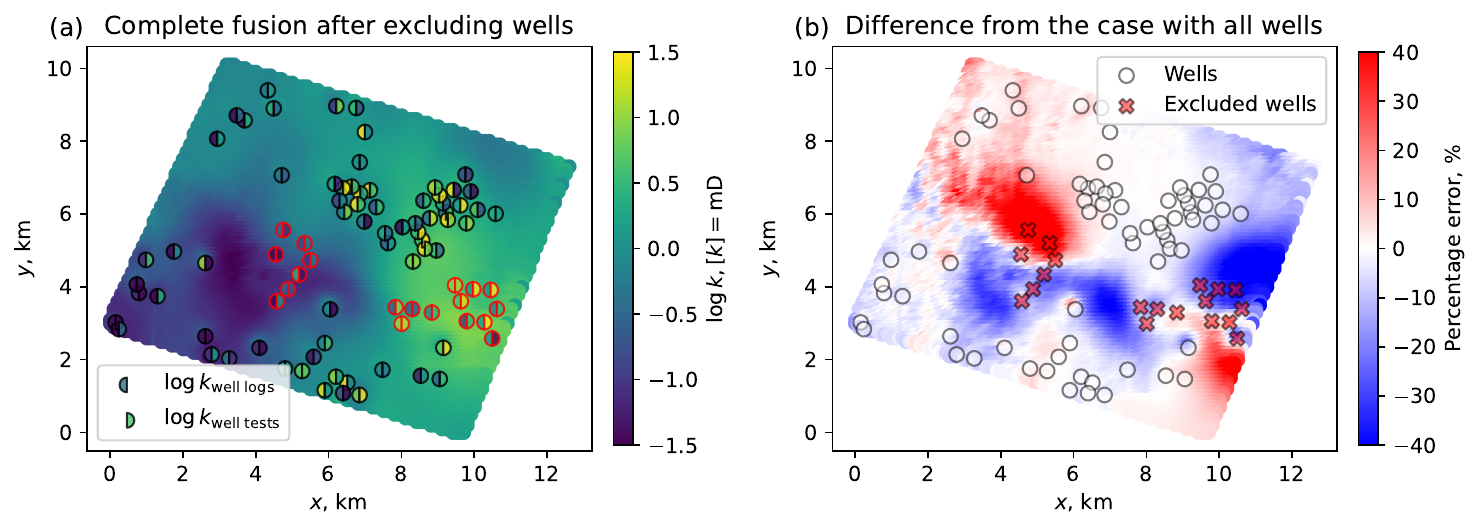}
\caption{Comparison of permeability maps: (a) new map, fused without excluded wells (marked in red), and (b) difference from the map fused with all wells, which was presented in Figure \ref{fig:complete_fusion}a.}
\label{fig:all_n_excluded}
\end{figure}

Employment of such an approach allowed the assessment of the differences in permeability predictions generated by the fusion model and facilitated the modeling of predictions in regions lacking direct permeability observations yet potentially promising for exploration. The comparison of permeability maps constructed with all wells included versus maps generated after excluding wells of high/low permeability is illustrated in Figure \ref{fig:all_n_excluded}.

In comparison to the "complete fusion" map that was fused with all wells with well logs and well tests (see Figure \ref{fig:complete_fusion}a), the new map fused after excluding wells (see Figure \ref{fig:all_n_excluded}a) noticeably differs in the regions of the removed wells. To analyze it, the percentage difference map was computed using the following formula. 
\begin{equation} \label{eq:percentage error2}
\text{Percentage error }(\vec{r}) = 100 \cdot \frac{ K_{\text{after excluding wells}}(\vec{r}) - K_{\text{with all wells}}(\vec{r})} {K_{\text{with all wells}} (\vec{r})}
\end{equation}
In the difference map, presented in Figure \ref{fig:all_n_excluded}b, red regions correspond to the areas where the permeability map fused after excluding wells has larger values than the one fused with all wells. Blue color, on the other hand, indicates regions in which the permeability became lower after removing wells. The white color corresponds to no change in permeability values. The difference map shows that in most regions near the remaining wells, the permeability values were not changed. In the regions where the wells were removed, either an increase or a decrease in permeability values is observed. This could happen due to two reasons. First, the information has propagated from the nearest remaining wells. If those wells have high or low interpreted permeability values, the regions will replicate those values. Second, the new seismic CNN has not learned too-high or too-low permeability values from excluded wells, and now its predictions are more moderate. This leads to an increase in permeability in regions that originally had low permeability and a decrease in permeability in regions that had high permeability values. 

The error metrics' values achieved for permeability predictions generated by the fusion model with seismic, “complete fusion”, and without seismic, “pure fusion”, are summarized in Table \ref{tab:metrics-complex}. The metrics include mean squared error (MSE) and coefficient of determination ($R^2$) along with their corresponding standard deviations. 

\begin{table}[h!!!]
\centering
\caption{Comparison of the performance of fusion model with and without seismic data, with all and excluded wells}
\label{tab:metrics-complex}
\resizebox{0.7\linewidth}{!}{
\begin{tabular}{c|cc|cc} 
\hline
                                         & \multicolumn{4}{c}{\textbf{Fusion}}                                                        \\ 
\hline
                                         & \multicolumn{2}{c|}{\textbf{Pure (without seismic)}}          & \multicolumn{2}{c}{\textbf{Complete (with seismic)}}    \\ 
\hline
\textbf{Metric}                   & \textbf{all}       & \textbf{excluded}      & \textbf{all}       & \textbf{excluded}       \\ 
\hline
\rowcolor[rgb]{0.937,0.937,0.937} MSE & 0.015 & 0.014 & 0.008 & 0.01 \\
$R^2$ & 0.96 & 0.917 & 0.972 & 0.935 \\
\hline
\end{tabular}
}
\end{table}

Based on Table \ref{tab:metrics-complex}, a noticeable reduction in the $R^2$ and an increase in MSE values were observed after excluding wells with high/low permeability for both the "pure fusion" and "complete fusion" models. This suggests that the ability of the fusion model to explain permeability variability decreased when fewer extreme-permeability data were available, potentially impacting the reliability of the predictions for areas with heterogeneous permeability. The complete fusion model (with seismic data) consistently exhibited lower MSE and higher R² compared to the "pure fusion" model, regardless of the inclusion or exclusion of the wells with extreme permeability values. This confirms that adding seismic data improves the reliability of permeability predictions generated by the fusion model. 

For all experiments, the optimized kernel values from Equations \eqref{bi_kernel}, \eqref{eq:well_log_kernel}, \eqref{eq: seismic kernel} are provided in Appendix A, Table \ref{tab:constant_opt_2}. 

%% file: 6_conclusion.tex
\section{Conclusions}
\label{sec:conclusions}

This paper presents the methodology of fusing geological data from well logs, pressure build-up tests, and seismic RMS amplitude cubes. The approach proposed uses kernel regression to fuse the data with a specially designed kernel shape, accounting for different data locality and reliability. A deep 3D-convolutional network is trained to predict permeability based on seismic data. Fused data from the first two sources, namely well log- and well test-based permeability estimations, helps to extend the effective number of training points for the seismic predictor and thus to prevent overfitting. Based on that,  the permeability map is rebuilt with the same kernel regression method, but including seismic information.

This approach is tested on permeability prediction for the part of a selected Western Siberia oilfield. The permeability map generated with seismic displays a better regression performance as well as more information in regions far from the wells: $MSE = 0.008$, $R^2 = 0.972$. Also, the map with seismic is validated by excluding wells in the low- and high-permeability zones, and the model successfully predicts the zones with excluded data. In addition, all maps were preprocessed with Q-Q transformation so that well logs show the same distribution as build-up test data, and this transformation alleviates the "bull's eye" effect.

The proposed methodology still needs several potential improvements to be considered in future studies. For instance, the incorporation of a broader range of seismic attributes and integrating them with well data directly within the neural network framework. Another direction that could be considered is to use prior reservoir geology knowledge by restricting the generated fields to certain patterns (e.g., to riverbeds for fluvial reservoirs). The approach can also be extended to the 3D case to account for multi-layeredness and fractures. Another potential extension involves conditioning on production data. The methodology could further be refined to incorporate initial data measurement errors and to produce the final results together with their uncertainties. Improving the validation procedure by considering metrics related to business goals is also a direction of further development (e.g., enhancing production forecast accuracy by using a fused permeability field as an initial input for model adaptation).

\newpage

\section*{List of abbreviations}

\begin{table}[h!!!!!!]
\begin{tabular}{|l|l|}
\hline
\textbf{Abbreviation} & \textbf{Definition} \\ \hline
2D, 3D & 2 dimensions, 3 dimensions \\ \hline
CNN & Convolutional neural network \\ \hline
IDW & Inverse distance weighting \\ \hline
LOOCV & Leave-one-out cross-validation \\ \hline
MSE & Mean squared error \\ \hline
PDF & Probability density function \\ \hline
Q-Q transformation & Quantile-quantile transformation \\ \hline
$R^2$ & Coefficient of determination \\ \hline
RMS amplitudes & Root mean square amplitudes \\ \hline
SCAL & Special core analysis \\ \hline
SVM & Support vector machine \\ \hline
WL & Well logs \\ \hline
WT & Well tests \\ \hline
\end{tabular}
\end{table}

\section*{CRediT authorship contribution statement}
\textbf{Grigoriy Shutov:} 
Writing - Original Draft,  Investigation, Software, Writing - Review \& Editing, Formal analysis, Visualization.
\textbf{Viktor Duplyakov:} Writing - Original Draft,  Investigation, Software, Writing - Review \& Editing, Formal analysis, Visualization.
\textbf{Shadfar Davoodi:} Writing - Original Draft, Visualization, Formal analysis, Writing - Review \& Editing
\textbf{Anton Morozov:} Writing - Original Draft, Investigation.
\textbf{Dmitriy Popkov:} Writing - Original Draft, Visualization.
\textbf{Kirill Pavlenko:} Writing - Original Draft, Investigation.
\textbf{Albert Vainshtein:} Supervision, Project administration.
\textbf{Viktor Kotezhekov:} Resources, Data Curation, Supervision.
\textbf{Sergey Kaygorodov:} Resources, Data Curation.
\textbf{Boris Belozerov:} Supervision, Validation, Project Administration, Funding acquisition.
\textbf{Mars M Khasanov:} Supervision, Project Administration, Funding acquisition.
\textbf{Vladimir Vanovskiy:} Conceptualization, Methodology, Writing - Review \& Editing, Project administration.
\textbf{Andrei Osiptsov:} Writing - Review \& Editing, Supervision, Funding acquisition.
\textbf{Evgeny Burnaev:} Supervision, Funding acquisition.

\section*{Funding}
This work was supported by the Ministry of Economic Development of the Russian Federation (code 25-139-66879-1-0003).

\section*{Acknowledgements}
The authors gratefully appreciate insightful discussions with experts from ``LLC Gazpromneft-STC'' and thank the Association ``Artificial intelligence in industry'' for providing a platform for such discussions.

%% file: 7_appendix.tex
\section{}

\subsection{Kernel values}

For the ablation study (see Section \ref{sec:ablation}), we provide the values of the kernel parameters, described by Equations \eqref{bi_kernel}, \eqref{eq:well_log_kernel}, \eqref{eq: seismic kernel}. They were optimized using Algorithm \ref{alg:kernel_opt}.

\begin{table}[h!!!]
\centering
\caption{Constants used for optimization \& optimal values for experiment with all wells and excluded areas}
\label{tab:constant_opt_2}
\resizebox{0.7\linewidth}{!}{%
\begin{tabular}{c|cc|cc|cc} 
\hline
\multicolumn{3}{c|}{}                                                             & \multicolumn{4}{c}{\textbf{Fusion}}                                                \\ 
\hline
\multicolumn{1}{c}{}                       & \multicolumn{2}{c|}{\textbf{Bounds}} & \multicolumn{2}{c|}{\textbf{Pure (without seismic)}}    & \multicolumn{2}{c}{\textbf{Complete (with seismic)}}  \\ 
\hline
\textbf{Parameter}                         & \textbf{min} & \textbf{max}          & \textbf{all} & \textbf{excluded}      & \textbf{all} & \textbf{excluded}           \\ 
\hline
\rowcolor[rgb]{0.937,0.937,0.937} $r_d$ & 100 & 300 & 296.5 & 294.1 & 299.0 & 299.7 \\
$r_g$ & 5 & 50 & 15.6 & 12.7 & 28.8 & 28.9 \\
\rowcolor[rgb]{0.937,0.937,0.937} $\alpha$ & 0.5 & 2 & 1.98 & 2.00 & 1.99 & 1.95 \\
$\beta$ & 1 & 2 & 1.00 & 1.02 & 1.00 & 1.00 \\
\rowcolor[rgb]{0.937,0.937,0.937} $\gamma$ & 0.01 & 2 & 0.12 & 1.13 & 0.27 & 0.04 \\
$\delta$ & 0.05 & 1 & 0.73 & 0.30 & 0.99 & 0.46 \\
\rowcolor[rgb]{0.937,0.937,0.937} $w_s$ & 0.1 & 0.5 & - & - & 0.42 & 0.25 \\ \hline
Metric & \multicolumn{2}{l|}{} & \multicolumn{2}{l|}{} & \multicolumn{2}{l}{}  \\ \hline
\rowcolor[rgb]{0.937,0.937,0.937} MSE &  &  & 0.015 & 0.014 & 0.008 & 0.01 \\
$R^2$ &  &  & 0.96 & 0.917 & 0.972 & 0.935 \\
\hline
\end{tabular}
}
\end{table}